\documentclass[12pt]{article}
\usepackage{epsfig}
\usepackage{amssymb,amsmath}
\usepackage{graphicx}

\newcommand{\bea}{\begin{eqnarray}}
\newcommand{\eea}{\end{eqnarray}}
 \makeatletter
\newcommand{\be}{\begin{equation}}
\newcommand{\ee}{\end{equation}}

\begin{document}
%
\vspace*{1.0cm}

\begin{center}
\baselineskip 20pt 
{\Large\bf 
Sparticle spectroscopy of the minimal SO(10) model
}
\vspace{1cm}

{\large 
Takeshi Fukuyama$^{~a,}$\footnote{ E-mail: fukuyama@se.ritsumei.ac.jp}, 
Nobuchika Okada$^{~b,}$\footnote{ E-mail: okadan@ua.edu}, 
and 
Hieu Minh Tran$^{~c,}$\footnote{ E-mail: hieu.tranminh@hust.edu.vn} 
}
\vspace{.5cm}

{\baselineskip 20pt \it
$^a$Research Center for Nuclear Physics (RCNP), \\ Osaka University, Ibaraki, Osaka, 567-0047, Japan\\
$^b$Department of Physics and Astronomy, University of Alabama, Tuscaloosa, AL35487, USA\\
$^c$Hanoi University of Science and Technology, 1 Dai Co Viet Road, Hanoi, Vietnam 
} 

\vspace{.5cm}

\vspace{1.5cm} {\bf Abstract}
\end{center}
The supersymmetric (SUSY) minimal SO(10) model is a well-motivated grand unified theory, 
  where the Standard Model (SM) fermions have Yukawa couplings 
  with only one ${\bf 10}$-plet and one $\overline{\bf 126}$-plet Higgs fields 
  and it is highly non-trivial if the realistic quark and lepton mass matrices 
  can be reproduced in this context. 
It has been known that the best fit for all the SM fermion mass matrices 
  is achieved by a vacuum expectation value of the $\overline{\bf 126}$-plet Higgs field 
  being at the intermediate scale of around ${\cal O}(10^{13})$ GeV. 
Under the presence of the SO(10) symmetry breaking at the intermediate scale, 
   the successful SM gauge coupling unification is at risk and likely to be spoiled. 
Recently, it has been shown that the low-energy fermion mass matrices, 
  except for the down-quark mass predicted to be too low, 
  are very well-fitted without the intermediate scale. 
In order to resolve the too-low down quark mass while keeping the other fittings intact, 
   we consider SUSY threshold corrections to reproduce the right down quark mass. 
It turns out that this requires flavor-dependent soft parameters. 
Motivated by this fact, we calculate particle mass spectra at low energies 
  with flavor-dependent sfermion masses at the grand unification scale. 
We present a benchmark particle mass spectrum which satisfies a variety of 
  phenomenological constraints, in particular, the observed SM-like Higgs boson mass of around 125 GeV 
  and the relic abundance of the neutralino dark matter 
  as well as the experimental result of the muon anomalous magnetic moment.  
In the resultant mass spectrum, sleptons in the first and second generations, 
  bino and winos are all light, and this scenario can be tested at the LHC Run-2 in the near future.

\thispagestyle{empty}

\newpage

\addtocounter{page}{-1}

\baselineskip 18pt

\section{Introduction} 

Although the supersymmetric (SUSY) extension of the Standard Model (SM) is one of the most promising candidates 
  for new physics beyond the SM, the results of the search for SUSY particles at the Large Hadron Collider (LHC) \cite{LHC} 
  has been increasing tensions to the minimal SUSY SM (MSSM).   
However, in most of MSSM analysis, the universal boundary conditions are taken into account
   for sfermion masses and gaugino masses at the grand unified theory (GUT) scale ($M_{GUT} \simeq 10^{16}$ GeV), 
   which seems to make the tensions stronger. 
It is worth considering a more general spectrum for sparticles \cite{SMass}.

Despite the tensions to the MSSM, the SUSY GUT is still a very attractive paradigm
  for physics beyond the SM at high energies, supported by the successful gauge coupling unification 
  at $M_{GUT} \simeq 10^{16}$ GeV in the context of the MSSM. 
Among various SUSY GUT models, the renormalizable minimal SUSY SO(10) model (minimal SO(10) model) has attracted a lot of attention, 
  where one ${\bf 10}$-plet and one $\overline{\bf 126}$-plet Higgs fields couple with ${\bf 16}$-plet MSSM matters.
The guiding principles for this GUT model are the minimality of the Higgs multiplets coupled with the MSSM matters 
  and the renormalizablity. 
The minimal SO(10) model can not only reproduce  the quark and lepton mass matrices at low energies 
  but also provide the symmetry breaking pattern from SO(10) to the SM gauge group almost unambiguously 
  \cite{Babu, Matsuda, FO, Clark, Aulakh, Bajc, BM}.
However, there has been a very serious tension to the model, namely, 
  in order to fit all the fermion mass matrix data at low energies  
  the $\overline{\bf 126}$-plet Higgs field needs to develop a vacuum expectation value (VEV) 
  at the intermediate scale of around $10^{13}$ GeV \cite{BM, Bertolini}. 
This indicates that many exotic states appear at the intermediate scale, 
  so that the success of the SM gauge coupling unification at the GUT scale is spoiled.

Recently, one of the authors (T.F.), together with K.~Ichikawa and Y.~Mimura, has revisited 
   the minimal SO(10) model to fit the fermion mass matrix data with a new strategy \cite{FIM}. 
Here, the authors of this paper have noticed that thanks to the recent progress in the neutrino oscillation experiments, 
  the mass matrix data in the lepton sector are more accurate than those in the quark sector. 
Instead of the quark mass matrix data, they have employed the mass matrix data in the lepton sector 
  as the inputs to fit the fermion mass matrices. 
Through the $\chi^2$-analysis for the general parameter space, they have found parameter regions 
  to nicely fit the fermion mass matrix data.  
Although the best fit is obtained with the intermediate scale VEV of the $\overline{\bf 126}$-plet Higgs field,  
  which is consistent with the result in \cite{BM, Bertolini}, 
  it has been found that a parameter set without the intermediate scale can results in a very good fit 
  for the fermion mass matrix data, except for the down quark mass being too small.

The fitting result with no intermediate scale is particularly interesting 
   since in this case the successful gauge coupling unification is kept intact. 
As discussed in \cite{FIM}, the too small down quark mass can be resolved 
   by SUSY threshold corrections through $A$-terms \cite{Hall}. 
However, not to alter the other fittings which are very good, the $A$-term corrections  
  must be sizable only for the down quark Yukawa coupling, in other words, 
  the soft SUSY breaking parameters are flavor-dependent.  
Motivated by this fact, we investigate a flavor-dependent sparticle mass spectrum 
   in the context of the minimal SO(10) model and a variety of phenomenological constraints. 
Considering very severe SUSY FCNC constraints, 
   we impose the universal mass for the sfermions in the matter ${\bf 16}$-plets   
   of the first and second generations. 
Among other phenomenological constraints, we focus on a parameter set 
   which can not only reproduce the measured SM-like Higgs boson of around 125 GeV and  
   the observed relic density of the neutralino dark matter, 
   but also fill the discrepancy between the experimental result and the SM prediction 
   for the muon anomalous magnetic dipole moment $a_\mu = \frac{1}{2} (g_\mu-2)$.  
In order for $a_\mu$ to receive sizable sparticle contributions, the universal mass inputs for 
   sfermions in the first and second generation ${\bf 16}$-plets as well as the gauginos must be $\lesssim 1$ TeV, 
   while the 125 GeV Higgs mass requires the soft mass input for the 3rd generation ${\bf 16}$-plet 
   to be much larger.  
Hence, we have a large mass splitting between sfermions in the first two generations and the third generation. 
A similar direction was previously addressed in \cite{Ibe:2013oha},
though our arguments have the unique reasoning for them.

\section{Minimal SO(10) model and fermion mass matrices }
In this section, we briefly review the minimal SO(10) model 
   and its fitting for the fermion mass matrices with the new strategy introduced in \cite{FIM},  
   focusing on the points which are used in this paper.

One of the most important features of the minimal SO(10) model is that all SM matter particles 
   are set in a single ${\bf 16}$-plet ($\psi$; $i=1,2,3$) for each generation. 
Renormalizability indicates that its coupling with the Higgs multiplets is limited in the Yukawa coupling: 
    ${\bf 16}\otimes {\bf 16}={\bf 10}\oplus {\bf 120}\oplus {\bf 126}$.
Therefore, the Higgs multiplet must belong to ${\bf 10}$, ${\bf 120}$ and ${\overline{\bf 126}}$ 
   to make the Yukawa couplings invariant under SO(10). 
One ${\bf 10}$-plet Higgs ($H$) is indispensable but not sufficient 
  since we have Cabbibo-Kobayashi-Maskwa (CKM) and Maki-Nakagawa-Sakata (MNS) mixing matrices. 
Minimally, we add only one ${\overline{\bf 126}}$-plet Higgs ($\overline{\Delta}$). 
This ${\overline{\bf 126}}$-plet Higgs (not ${\bf 120}$-plet Higgs) is required from the following reasons. 
Under the Pati-Salam subgroup of SO(10), the ${\overline{\bf 126}}$-plet Higgs includes 
   both SU(2)$_R$ and SU(2)$_L$ triplets ($({\overline{\bf 10}},{\bf 1}, {\bf 3})$ and $({\bf 10}, {\bf 3}, {\bf 1}))$
   which generate the neutrino mass matrix 
   through the type I~\cite{typeI} and type II~\cite{typeII} seesaw mechanism, respectively. 
Also the VEVs of the ${\bf126}+\overline{\bf 126}$-plet Higgs fields in the $B-L=\pm 2$ direction  
   reduce the rank of the SO(10) gauge symmetry in the renormalizable model 
   and guarantees the conservation of $R$-parity. 
Thus, in the minimal SO(10) model the Yukawa coupling has the form of 
\begin{equation}
W_Y= \frac12  {\bf h}_{ij} \psi_i \psi_j H + \frac12  {\bf f}_{ij} \psi_i \psi_j \bar \Delta.
\end{equation}
Due to the SO(10) algebra, the coupling matrices are symmetric, ${\bf h}_{ij} = {\bf h}_{ji}$ and ${\bf f}_{ij} = {\bf f}_{ji}$.

After the SO(10) symmetry is broken down to the SM gauge symmetry, the fermion Yukawa matrices are given as follows:
\begin{eqnarray}
&& Y_u = h + r_2 f, \nonumber \\
&& Y_d = r_1 (h + f),  \nonumber \\
&& Y_e = r_1 (h - 3 f), \nonumber \\
&& Y_\nu = h- 3r_2 f,  
\label{massformula}
\end{eqnarray}
where $r_1$ and $r_2$ depend on the Higgs mixing (doublet Higgs mixing in {\bf 10} and $\overline{\bf 126}$),
and 
$h$ and $f$ are the original Yukawa matrices $\bf h$ and $\bf f$ multiplied by Higgs mixings,
\begin{eqnarray}
h = V_{11} {\bf h}, \qquad
f = \frac{U_{12}}{\sqrt3 r_1} {\bf f}, 
\qquad
r_1 = \frac{U_{11}}{V_{11}},
\qquad
r_2 = r_1 \frac{V_{13}}{U_{12}}, 
\end{eqnarray}
where $U$ and $V$ are the unitary matrices to diagonalize the mass matrices 
  of the MSSM Higgs doublets in the ${\bf 10}$-plet and $\overline{\bf 126}$-plet. 
See \cite{Dutta:2005ni} for details.

The charged fermion mass matrix and the Dirac neutrino mass matrix are obtained as 
\begin{equation}
M_u = Y_u v_u, \qquad
M_d = Y_d v_d, \qquad
M_e = Y_e v_d, \qquad
M_\nu^D = Y_\nu v_u.
\end{equation}
where $v_u$ and $v_d$ are
the VEVs of up- and down-type Higgs fields. From the mass formula Eq.~(\ref{massformula}), 
  we obtain the relation of the mass matrices as
\begin{eqnarray}
&& M_d = M_e + \frac{4}{1-r_2} F = r M_u + F, \\
&& r M_\nu^D = M_e + 3F,
\end{eqnarray}
where 
\begin{equation}
r = r_1 \frac{v_d}{v_u} \equiv r_1 \cot\beta,
\end{equation}
and the matrix $F$ is
\begin{equation}
F = r(1-r_2) f v_u.
\end{equation}
Roughly, we obtain $r \sim m_b/m_t$.

The right-handed Majorana neutrino mass matrix is obtained
as
\begin{equation}
M_R = \sqrt2 {\bf f} v_R,
\end{equation}
where $v_R$ is a VEV of $\overline{\bf 126}$.
Practically, we denote
\begin{equation}
M_R = c_R v_R f,
\end{equation}
where 
\begin{equation}
c_R = \sqrt6 \frac{r_1}{U_{12}} = \sqrt6 \frac{r_2}{V_{13}} 
\end{equation}
 for the current notation.
Because $U_{12}$ and $V_{13}$ are components of the diagonalization unitary matrix for doublet Higgs fields,
$c_R$ has a minimal value.
The size of $c_R$ is related to the size of original ${\bf f}$ coupling,
which will be important to derive the GUT scale threshold correction
for flavor violation. 
%
%
The seesaw neutrino mass matrix can be written as \cite{typeI, typeII} 
\begin{equation}
{\cal M}_\nu = M_L - M_\nu^D M_R^{-1} (M_\nu^D)^T,
\end{equation}
where
$M_L$ is the left-handed neutrino Majorana mass which comes from $SU(2)_L$ triplet coupling,  
\begin{equation}
M_L = c_L v_L f.
\end{equation}

We note here that the SO(10) breaking vacuum can be specified
  by one complex parameter for the minimal Higgs contents:
  ${\bf 10}+{\bf 126}+\overline{\bf 126}+{\bf 210}$~\cite{Clark, Aulakh}. 
The ${\bf 210}$-plet is required to connect the ${\bf 10}$-plet and the $\overline{\bf 126}$-plet Higgs fields.
The coefficient $r_2$ is determined by the complex parameter. 
This is one of the reason why the model with the minimal number of Higgs multiplets in the Higgs superpotential 
  does not provide a solution consistent with the gauge coupling unification in the minimal SUSY version 
  of the minimal SO(10)~\cite{Bertolini}.  

In Ref.~\cite{FIM},  using the parameters in lepton sector as inputs, the $\chi^2$ analysis has been performed 
  for fitting the quark masses and mixings under the current observables.
The fit provides information about the behavior of the fitting parameters with respect to 
  the scale of ${\bf 126}+\overline{\bf 126}$ VEVs ($v_R$).
The scale plays a key role for the breaking the rank-5 SO(10) symmetry down to the rank-4 SM gauge symmetry 
  and is also important to see how the light neutrino mass is generated by the seesaw mechanism \cite{typeI, typeII}.

The values and the uncertainties used in the fit at the GUT scale are summarized in Table~\ref{table:parameters}, 
   where the quark mass values are taken from Table~III in Ref.~\cite{Bora:2012tx} (MSSM, $\tan \beta = 10$), 
   the quark mixing angles and the CP-phase at the GUT scale are evaluated 
   from the PDG data \cite{Agashe:2014kda} by $\overline{{\rm DR}}$ scheme, 
   and the neutrino oscillation data are taken from the global fit results in Ref.~\cite{Capozzi:2013csa}.

\begin{table}[t]
\begin{minipage}[t]{.45\textwidth}
\begin{tabular}{|c|c|c|}
\hline
\multicolumn{3}{|c|}{ Fixed Values } \\
\hline
$m_{u}$       & $ 3.961  \times 10^{-4} ~{\rm GeV} $ &  \cite{Bora:2012tx} \\
$m_{t}$       & $ 71.0883               ~{\rm GeV} $ &  \cite{Bora:2012tx} \\
$m_{e}$       & $ 3.585  \times 10^{-4} ~{\rm GeV} $ &  \cite{Bora:2012tx} \\
$m_{\mu}$     & $ 7.5639 \times 10^{-2} ~{\rm GeV} $ &  \cite{Bora:2012tx} \\
$m_{\tau}$    & $ 1.3146                ~{\rm GeV} $ &  \cite{Bora:2012tx} \\
$\Delta m_{\rm sol}^2$    & $ 7.54 \times 10^{-5} ~{\rm eV}^2 $   &  \cite{Capozzi:2013csa} \\
$\Delta m_{\rm atm}^2$    & $ 2.4  \times 10^{-3} ~{\rm eV}^2 $   &  \cite{Capozzi:2013csa} \\
$\theta_{12}$    &    0.583           &  \cite{Capozzi:2013csa} \\
$\theta_{23}$    &    0.710, $\pi/4$  &  \cite{Capozzi:2013csa} \\
$\theta_{13}$    &    0.156           &  \cite{Capozzi:2013csa} \\
\hline
\end{tabular}
\end{minipage}
\begin{minipage}[t]{.45\textwidth}
\begin{tabular}{|c|c|c|}
\hline
\multicolumn{3}{|c|}{ Parameters to fit } \\
\hline
$m_c$    & $ 0.1930  \pm 0.025              ~{\rm GeV} $   &    \cite{Bora:2012tx}    \\
$m_d$    & $ (9.316   \pm 3.8)   \times 10^{-4} ~{\rm GeV} $    &    \cite{Bora:2012tx}    \\
$m_s$    & $ (1.76702 \pm 0.5)   \times 10^{-2} ~{\rm GeV} $    &   \cite{Bora:2012tx}    \\
$m_b$    & $ (0.9898  \pm 0.03)                 ~{\rm GeV} $    &    \cite{Bora:2012tx}    \\
$     V_{us}$    &    $ 0.224 \pm 0.002 $ &    \cite{Agashe:2014kda}    \\
$     V_{cb}$    &    $ (3.7 \pm 0.13)  \times 10^{-2} $    &  \cite{Agashe:2014kda} \\
$     V_{ub}$    &    $ (3.7 \pm 0.45)  \times 10^{-3} $    &  \cite{Agashe:2014kda} \\
$\delta_{\mathrm{KM}}$    &   $ 1.18 \pm  0.2 $    &  \cite{Agashe:2014kda}  \\
\hline
\end{tabular}
\end{minipage}
\caption{\sl \small 
The reference parameters used in the fit.
}
\label{table:parameters}
\end{table}

In Ref.~\cite{FIM}, the best fit is obtained when $v_R \simeq 10^{13}$ GeV as was reported in Refs.~\cite{BM, Bertolini}.
However, such a low VEV scale spoils the successful gauge coupling unification 
in the context of the MSSM~\cite{BM, Bertolini}.
Therefore, we focus on solutions for $v_R$ being around the GUT scale. 
In Table~\ref{table:fit}, we cite the results in Ref.~\cite{FIM} for the type II seesaw dominated case 
  for the neutrino mass matrix. 
Although they are not the best fit found in Ref.~\cite{FIM}, the fits are very good 
  except for the down quark mass. 
 

\begin{table}[p]
\begin{center}
\begin{tabular}{|c|c|c|}
\hline
\multicolumn{3}{|c|}{ Type II } \\
\hline
\hline
$ c_Rv_R~({\rm GeV})   $ & $ 8.86  \times 10^{16} $ &$ 9.22  \times 10^{16} $ \\ \hline
$\theta_{23}$         & $  0.710    $&$  \pi/4    $\\ \hline
$\alpha_{e}$          & $ -0.66648  $&$ -0.25301  $\\
$\alpha_{\mu}$      &$ -2.8148   $&$  2.8177   $\\
$\alpha_{\tau}$      &$ -0.53961  $&$ -0.84287  $\\
$\alpha_{2}$          &$ -2.8709   $&$ -3.1146   $\\
$\alpha_{3}$          &$ -1.9809   $&$ -2.8604   $\\
$\delta_{{\rm PMNS}}$          &$ -2.3550   $&$ -3.1131   $\\
$\log_{10} (m_{1}/{\rm GeV})$   &$-11.207    $&$-11.173    $\\
$|\delta|$                    &$ 15.545    $&$ 16.156    $\\
${\rm Arg(\delta)}$           &$  0.43912  $&$  0.51567  $\\ \hline
$m_{c}~({\rm GeV})$       &$  0.1978   $&$  0.1989   $\\
$m_{d}~({\rm GeV})$       &$  0.0004138$&$  0.0003936$\\
$m_{s}~({\rm GeV})$       &$  0.01980  $&$  0.02028  $\\
$m_{b}~({\rm GeV})$       &$  0.9903   $&$  0.9901   $\\
$     V_{uc}$                  &$  0.2240   $&$  0.2241   $\\
$     V_{sb}$                  &$  0.003765 $&$  0.003724 $\\
$     V_{ub}$                  &$  0.03694  $&$  0.03695  $\\
$\delta_{{\rm KM  }}$     &$  1.195    $&$  1.160    $\\ \hline
\multicolumn{3}{|c|}{Pull} \\ \hline
$m_{c}$               &$    0.191 $&$    0.236 $\\
$m_{d}$               &$   -1.363 $&$   -1.416 $\\
$m_{s}$               &$    0.426 $&$    0.522 $\\
$m_{b}$               &$    0.017 $&$    0.010 $\\
$     V_{uc}$         &$   -0.000 $&$    0.027 $\\
$     V_{sb}$         &$    0.144 $&$    0.054 $\\
$     V_{ub}$         &$   -0.044 $&$   -0.039 $\\
$\delta_{{\rm KM}}$   &$    0.075 $&$   -0.099 $\\
\hline
$ r               $      & $ 0.0230         $&$ 0.0233 $\\
$ r_{2}           $    & $ 2.15 + 0.227 i $&$ 2.22 - 0.160 i  $\\
$\chi^2$              &$ 2.10  $&$ 2.35   $\\
\hline
\end{tabular}
\caption{\sl \small 
The fit result for type II (which is called Type I+II in many literatures) cited from Ref.~\cite{FIM}. }
\label{table:fit}
\end{center}
\end{table}

In the results, the main pull comes from the down quark mass which is too low. 
However, this small down quark mass is advantageous for making the proton lifetime 
  longer under the relation of $({\cal M}_\nu)_{11}=({\cal M}_\nu)_{12}=0$ \cite{Dutta:2005ni, FIM2}.
In order to resolve the too-low down quark mass, 
  we consider SUSY threshold corrections from the gluino loop diagrams \cite{Hall} as  
\begin{equation}
\delta m_d= \frac{\alpha_s}{3\pi}\frac{m_{\tilde{g}}}{m_{\tilde q}^2} \left( m_d \mu\tan\beta+ \frac{v_d a_d}{\sqrt{2}} \right),  
\label{threshold}
\end{equation}
where $m_{\tilde g}$ and $m_{\tilde q}$ are the gluino and squark masses, respectively, 
 $v_d$ is the VEV of the down-type Higgs doublet, and $a_d$ is the $A$-term parameter defined as 
\begin{equation}
{\cal L}_{\rm soft} \supset  a_d \tilde{d}^\dagger \tilde{Q} H_d+ h.c. 
\end{equation}
   in the trilinear scalar coupling among the squarks and the down-type Higgs doublet.  
For example, when we take $m_{\tilde g} \sim m_{\tilde q} \sim 1$ TeV,  
  $a_d\sim 10$ GeV is sufficient to yield the right size of the down quark mass 
  from the SUSY threshold corrections.  
In analysis of sparticle mass spectrum, a universal $A$-term parameter ($A_0$) is often used, 
  which has a relation to $a_d$ as $Y_d A_0=a_d$ with $Y_d$ being a down-type quark Yukawa coupling. 
We can see that if we apply the universal $A$-term parameter for the SUSY threshold corrections, 
  the strange and bottom quarks also receive sizable corrections which ruin the good mass fitting. 
Hence, non-universal $A$-term parameters are necessary.

\section{Sparticle mass spectroscopy and current  experiments}
In the minimal SO(10) model, the quark and lepton supermultiplets in a same generation are unified 
  into a ${\bf 16}$-representation. 
Therefore, their corresponding soft masses and trilinear couplings are also unified in the same way,  
   significantly reducing the total number of free parameters compared to the general MSSM case.
For simplicity, we assume that at the GUT scale the soft masses of the first and the second generations 
  are universal while that of the third generation is independent.
As a result, in our analysis, the boundary conditions for the soft parameters at the GUT scale 
  consist of five free parameters  $\{M_{1/2}, m_0, m_3, A_0, \tan \beta \}$ 
which are, respectively, the universal gaugino masses, 
  the universal masses among the first two generations sfermions and the Higgs doublets, 
  the third generation sfermion mass, the universal trilinear coupling,
  and the ratio between the VEVs of two MSSM Higgs doublets.\footnote{
As we have discussed, non-universal trilinear couplings are essential to resolve 
  the too-low down quark mass while keeping the other fermion masses intact. 
Nevertheless, we can employ the universal trilinear coupling in our analysis 
   for the renormalization group (RG) evolutions of the soft parameters, 
  since the trilinear couplings for the first two generations contribute to the RG evolutions 
  only through their Yukawa couplings and their effects are negligibly small. 
}

For numerical analysis, we employ SOFTSUSY version 3.7.3 \cite{Allanach:2001kg} 
   to solve the RG equations with inputs at the GUT scale, and calculate 
   the corresponding sparticle mass spectrum at the low energies. 
The constrained observables are computed 
   by micrOMEGAs version 4.3.1 \cite{Belanger:2014vza,Belanger:2004yn,Belanger:2001fz}.
The Higgs sector is tested with the package HiggsBounds version 4.3.1
\cite{Bechtle:2008jh,Bechtle:2011sb,Bechtle:2013gu,Bechtle:2013wla}
   by considering the exclusion limits from Higgs searches at the LEP, the Tevatron and the LHC experiments. 
We also check the validity of the benchmark with the package SModelS version 1.0.4 \cite{Kraml:2013mwa} 
  which decomposes the model signal into simplified model spectrum topologies and compares with the LHC bounds.

\begin{table}
\label{benchmarks}
\begin{center}
\scalebox{1}[1]{
\begin{math}
\begin{array}{|c|c|}
\hline

   M_{1/2} &       900  \\

       m_0 &       393.515  \\

     m_3 &         9000  \\

       A_0 &       -13920  \\

\tan \beta &	10			 \\
\hline \hline

         h &     125.09  \\

  H^0, A^0 & 	8682  \\

     H^\pm &       8683  \\

 \tilde{g} &       2070  \\

\tilde{\chi}^0_{1,2}   &   399.7, 768.4  \\

\tilde{\chi}^0_{3,4}   &   8739, 8740  \\

\tilde{\chi}^\pm_{1,2} &   768.6 , 8740  \\

\tilde{u}, \tilde{c}_L &       1558  \\

\tilde{u}, \tilde{c}_R &       1524  \\

\tilde{d}, \tilde{s}_L &       1560  \\

\tilde{d}, \tilde{s}_R &       1514  \\

\tilde{t}_{1,2} &   5832, 7649  \\

\tilde{b}_{1,2} &   7644, 9080  \\

\tilde{\nu}^{e,\mu}_L &       585.3, 583.8     \\

\tilde{e}, \tilde{\mu}_L &       590.6, 589.1    \\

\tilde{e}, \tilde{\mu}_R &       404.0, 399.8   \\

\tilde{\nu}^\tau_L &       8976  \\

\tilde{\tau}_{1,2} &  8903, 8978    \\

\hline

\Delta \rho		&	3.30 \times 10^{-7}	\\
\Delta a_\mu 	&	1.39 \times 10^{-9}	\\
\text{BR}(b \rightarrow s \gamma)	&	3.33 \times 10^{-4}	\\
\text{BR}(B^0_s \rightarrow \mu^+ \mu^-)	&	3.08 \times 10^{-9}	\\
\frac{\text{BR}^{MSSM}(B^\pm \rightarrow \tau^\pm \nu_\tau)}{\text{BR}^{SM}(B^\pm \rightarrow \tau^\pm \nu_\tau)}	&	1.00	\\
\text{BR}(D^\pm_s \rightarrow \tau^\pm \nu_\tau) &	5.17 \times 10^{-2}	\\
\text{BR}(D^\pm_s \rightarrow \mu^\pm \nu_\mu) &	5.33 \times 10^{-3}	\\
R_{l23}		&	1.000	\\
\Omega h^2	&	0.1188	\\
\sigma_\text{SI}^{\chi-p} ({\rm pb})	&	1.507 \times 10^{-13}	\\
\sigma_\text{SD}^{\chi-p}	({\rm pb}) &	4.589 \times 10^{-8}	\\
\hline
\end{array}  
\end{math}}
\caption{
Benchmark particle mass spectrum for a set of inputs (in units of GeV).
$M_{1/2}, m_0, m_3, A_0$ are, respectively, the universal gaugino mass, 
the universal sfermion soft mass of the 1st and 2nd generations and Higgs doublets, 
the soft mass of the 3rd generation sfermions, and the common trilinear coupling at the GUT scale.
}
\label{table:BM}
\end{center}
\end{table} 

We consider a variety of phenomenological constraints from collider physics.
While the LEP limits is important to constrain the sleptons, neutralino and chargino sectors, the LHC experiments 
   are more sensitive to colored sparticles. 
The SM-like Higgs boson mass obtained by the combined ATLAS and CMS analysis \cite{Aad:2015zhl},
   the sparticle contributions ($\Delta \rho$) to the electroweak $\rho$-parameter \cite{Agashe:2014kda}, 
   and the anomalous muon magnetic moment~\cite{Bennett:2006fi, Roberts:2010cj, Hagiwara:2011af, Aoyama:2012wk} are taken into account.
Other constraints are from the branching ratios of rare decay processes:    
$b \rightarrow s + \gamma$ \cite{hfag},
$B^0_s \rightarrow \mu^+ \mu^-$	\cite{Olive:2016xmw},
$B^\pm \rightarrow \tau^\pm \nu_\tau$ \cite{hfag},
$D^\pm_s \rightarrow \tau^\pm \nu_\tau$	\cite{Agashe:2014kda},
$D^\pm_s \rightarrow \mu^\pm \nu_\mu$	\cite{Agashe:2014kda},
and a constraint on the Kaon system characterized by the decay parameter \cite{Antonelli:2008jg}
\begin{equation}
R_{l23} = \left|	\frac{V_{us}(K_{l2})}{V_{us}(K_{l3})} \times
			\frac{V_{ud}(0^+ \rightarrow 0^+)}{V_{ud}(\pi_{l2})} \right|,
\end{equation}
where the CKM matrix elements, $V_{us}$ and $V_{ud}$, are measured 
  from the corresponding 3-body semileptonic Kaon decay ($K_{l3}$), 
  2-body leptonic Kaon and pion decay ($K_{l2}$, $\pi_{l2}$), 
  and super-allowed nuclear beta decay ($0^+ \rightarrow 0^+$).
The limits for these constraints are as follows
\begin{eqnarray}
m_h	=	125.09 \pm 0.21 \text{(stat.)} \pm 0.11 \text{(syst.)}	,	&&	\\
\Delta \rho < 8.8 \times 10^{-4}	,	&&	\\
7.73 \times 10^{-10} < \Delta a_\mu < 42.14 \times 10^{-10} ,	&&	\\
3.11 \times 10^{-4} < \text{BR}(b \rightarrow s + \gamma) < 3.87 \times 10^{-4}	,	& [2\sigma] &	\\
1.7 \times 10^{-9} < \text{BR}(B^0_s \rightarrow \mu^+ \mu^-) < 4.3 \times 10^{-9},		& [2\sigma] &	\\
0.70  < 
\frac{\text{BR}^{exp}(B^\pm \rightarrow \tau^\pm \nu_\tau)}{\text{BR}^{SM}(B^\pm \rightarrow \tau^\pm \nu_\tau)}
 < 1.82,	& [2\sigma] &	\\
5.07 \times 10^{-2} < \text{BR}(D^\pm_s \rightarrow \tau^\pm \nu_\tau) < 6.03 \times 10^{-2},	& [2\sigma] &	\\
5.06 \times 10^{-3} < \text{BR}(D^\pm_s \rightarrow \mu^\pm \nu_\mu) < 6.06 \times 10^{-3}, 	& [2\sigma] &	\\
R_{l23} = 1.004 \pm 0.007. &&
\end{eqnarray}

Assuming the $R$-parity conservation, the model predicts the lightest neutralino to be a dark matter (DM) candidate 
    that can explain a wide range of cosmological observations such as gravitational lensing and galaxy rotation curves.
Therefore, we furthermore take into account constraints from DM searches 
    which play an essential role in the SUSY phenomenology and DM physics.
The DM relic density is precisely measured by the Planck experiment
\cite{Ade:2015xua}:
\begin{eqnarray}
\Omega_\text{CDM} h^2 &=& 0.1188 \pm 0.0010	.
\end{eqnarray}
Regarding to DM direct detections, the most severe constraint on spin-independent cross section 
  between DM and nuclei is set by the LUX experiment \cite{Akerib:2015rjg}:
\begin{eqnarray}
\sigma_\text{SI}^{\chi-p} \lesssim 8 \times 10^{-10} \text{ pb},
	\qquad 		m_{\text{WIMP}} \approx 400 \text{ GeV}. 
\end{eqnarray}
On the other hand, the DM indirect detection at the IceCube experiment \cite{Aartsen:2016exj} 
   with the $\nu \bar{\nu}$ channel imposes the most stringent upper limits 
   on the spin-dependent cross section between DM and nuclei for a DM mass of about a few hundreds GeV:
\begin{eqnarray}
\sigma_\text{SD}^{\chi-p} \lesssim 10^{-5} \text{ pb},
	\qquad 		m_{\text{WIMP}} \approx 400 \text{ GeV}.
\end{eqnarray}

\begin{figure}[h]
\begin{center}
\includegraphics[scale=1.2]{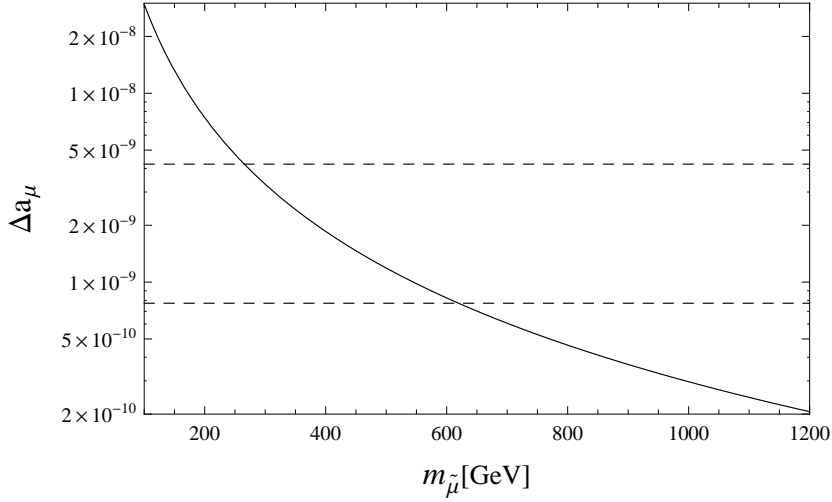}
  \caption{
The rough estimate of $\Delta a_\mu$ as a function of smuon mass, 
  along with the experimental result, $7.73 \times 10^{-10} < \Delta a_\mu < 42.14 \times 10^{-10}$
  (dashed horizontal lines). 
}\label{Fig:amu}
\end{center}  
\end{figure}

In Table \ref{table:BM}, a benchmark mass spectrum is shown for a given set of inputs at the GUT scale 
    with $\tan \beta = 10$ corresponding to the $\chi^2$ fitting results in Ref.~\cite{FIM}. 
The benchmark satisfies all the above phenomenological constraints. 
Here, the heavy squark masses and large trilinear couplings of the third generation at the order of $\mathcal{O}(10)$ TeV 
    play a crucial role to reproduce the Higgs boson mass of around 125 GeV via quantum corrections.%
\footnote{
The large $A_0$ in Table \ref{table:BM} may generate a true minimum away from the SM vacuum 
  and make the SM vacuum unstable. 
We have checked that the absolute stability conditions for the SM vacuum are satisfied in the direction 
  of the 3rd generation sfermions, namely, $A_t^2 < 3(m_{\tilde{t}_L}^2 + m_{\tilde{t}_R}^2 + m_{H_u}^2 + |\mu|^2 ) $ \cite{Camargo-Molina:2013sta}.
On the other hand, the stability conditions in the directions of the 1st and 2nd generation sfermions are not satisfied, 
  since they are relatively light compared to the corresponding $A$-term parameters.
However, such charge/color breaking vacua appear far away from the SM vacuum 
  because of the very small Yukawa couplings of the 1st and 2nd generations and 
  the tunneling rate is extremely small \cite{Kusenko:1996jn}.
Hence, we conclude that our benchmark spectrum is cosmologically safe.
}
Since the third generation sfermions are all heavy, their visible effect is only the contribution to the Higgs boson mass.
We find that the SUSY contribution to the muon anomalous magnetic moment is at the order of 
     $\mathcal{O}(10^{-9})$ which satisfies the experimental limits at the level of 2$\sigma$.
It is due to a large mass splitting between the first two generations and the third generation sfermions 
   and relatively light smuons and gauginos. 
In order to illustrate a response of $\Delta a_\mu$ to the sparticle mass, 
   we show $\Delta a_\mu$ as a function of smuon mass in Fig.~\ref{Fig:amu}. 
Here, we have used a very rough formula for the SUSY contributions as \cite{Moroi}
\bea 
  \Delta a_\mu \sim    \frac{\alpha_2}{4 \pi}  \left(\frac{m_\mu}{m_{\tilde \mu}}\right)^2\tan \beta, 
\eea 
where $\alpha_2$ is the SM SU(2) gauge coupling.

For this benchmark, the lightest sparticle (LSP) is a bino-like neutralino 
   playing the role of the DM, and the next-to-the LSP (NLSP) is the right-handed smuon.
A mass degeneracy between these LSP and NLSP is required to yield the correct DM relic density 
  via neutralino-slepton co-annihilation processes.
Since the scattering cross sections between the neutralino LSP and nuclei are a few orders of magnitude 
   smaller than the current sensitivity of the LUX experiment,  drastic improvements on new generations 
    of DM direct and indirect searches are necessary to capture signatures of the bino-like neutralino.
In spite of that, the light sleptons, neutralinos and charginos can be explored at the LHC Run-2 in the near future.
The projected high energy linear collider with the center of mass energy around 1 TeV can shed more lights 
    on the physics of these relatively low mass sparticles due to its high precision.

\begin{figure}[h]
\begin{center}
\includegraphics[scale=1.2]{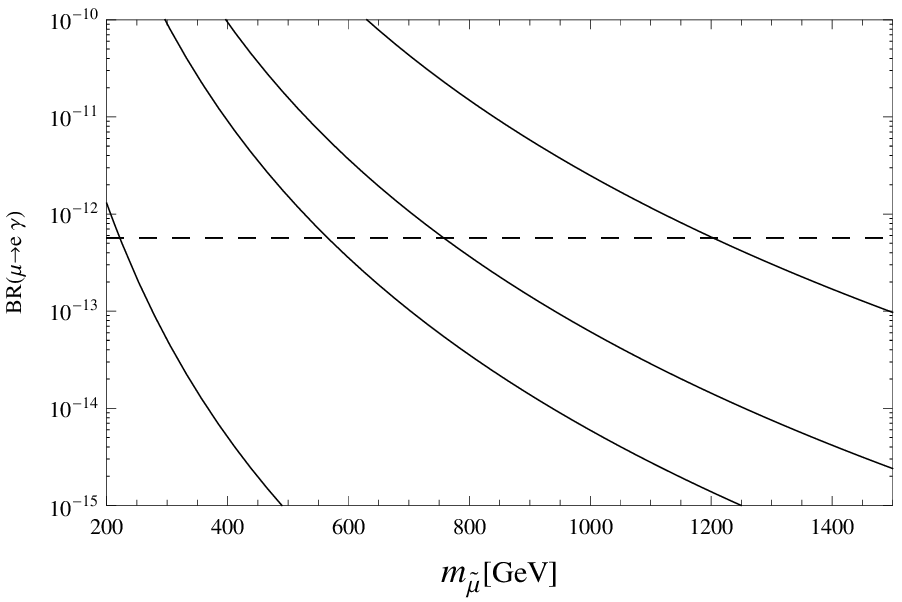}
\end{center}  
  \caption{
The branching ratio BR($\mu\to e \gamma$) as a function of the left-handed slepton mass. 
The solid lines from left to right correspond to a variety of choices of the GUT scale, 
  $M_{GUT}=1.20$, $1.23$, $1.30$, and $2.0$ in units of $10^{16}$ GeV, respectively. 
The current experimental upper bound by the MEG experiment~\cite{MEG} 
   is denoted as the horizontal dashed line. 
}\label{Fig:LFV}
\end{figure}

In the minimal SO(10) model, after the parameter fitting, 
  the neutrino Dirac Yukawa coupling ($Y_\nu$) is unambiguously determined. 
Using this neutrino Dirac Yukawa couplings, we calculate the lepton flavor violating (LFV) processes. 
The LFV effect most directly emerges in the left-handed slepton mass matrix 
   generated through the RG equations such as \cite{Hisano-etal}
\begin{eqnarray}
\mu \frac{d}{d \mu} 
  \left( m^2_{\tilde{\ell}} \right)_{ij}
&=&  \mu \frac{d}{d \mu} 
  \left( m^2_{\tilde{\ell}} \right)_{ij} \Big|_{\mbox{MSSM}} 
 \nonumber \\
&+& \frac{1}{16 \pi^2} 
\left( m^2_{\tilde{\ell}} Y_{\nu}^{\dagger} Y_{\nu}
 + Y_{\nu}^{\dagger} Y_{\nu} m^2_{\tilde{\ell}} 
 + 2  Y_{\nu}^{\dagger} m^2_{\tilde{\nu}} Y_{\nu}
 + 2 m_{H_u}^2 Y_{\nu}^{\dagger} Y_{\nu} \right)_{ij}  ,
 \label{slepton_RGE} 
\end{eqnarray}
where the first term in the right hand side denotes the normal MSSM term with no LFV.
We employ the data obtained in Ref.~\cite{FIM} for the heavy Majorana neutrino mass eigenvalues, 
\begin{equation}
M_{R_1} = 6.9 \times 10^8\ {\rm GeV},\quad
M_{R_2} = 3.3 \times 10^{14}\ {\rm GeV},\quad
M_{R_3} = 1.2 \times 10^{16}\ {\rm GeV},
\end{equation}
and the neutrino Dirac Yukawa coupling,  
\begin{equation}
Y_\nu
= \left(
 \begin{array}{ccc}
    0.000111 & 0.000203 + 0.000217\,i & 0.00888 + 0.00372\, i \\ 
    0.000440 & 0.0308 - 0.0248\, i & 0.0426+ 0.0013 \,i \\
    0.00607  & -0.0276 - 0.0069\, i & 0.990 - 0.278\, i
 \end{array}
\right),
\end{equation}
   for the case of $c_R v_R = 8.86 \times 10^{16}$ GeV (the second column in Table~\ref{table:fit}). 
We solve Eq.~(\ref{slepton_RGE}) from the GUT scale to low energies 
   with the inputs of the Majorana neutrino masses, the $Y_\nu$ data, 
   the slepton mass matrix $m_{\tilde \ell}=m_{\tilde \nu}={\rm diag}(m_0, m_0, m_3)$ 
   and $m_{H_u}=m_0$.  
Using the generated off-diagonal elements of the (left-handed) slepton mass matrix squared $m_{\tilde \ell}^2$  at low energies, 
   we roughly estimate the LFV decay rate of the charged leptons by \cite{Hisano-etal}
\bea
   \Gamma(\ell_i \to \ell_j \gamma ) \sim \frac{\alpha_{em}}{4} m_{\ell_i}^5  \times  \frac{\alpha_2^2}{4 \pi}  
    \frac{ |(\Delta m_{\ell}^2)_{ij}|^2}{m_{\tilde \mu_L}^8 } \tan^2 \beta .
\eea   
Our results for the branching ratio of the $\mu \to e \gamma$ process, which provides the most severe constraint, 
    are shown in Fig.~\ref{Fig:LFV} for a various values of the GUT scale, 
    along with the final results of the MEG experiment~\cite{MEG}, ${\rm BR}(\mu^+ \to e^+ \gamma) < 4.2 \times 10^{-13}$ 
    at 90\% C.L. (horizontal dashed line).    
We find that the resultant branching ratio is sensitive to the choice of $M_{GUT}$, 
   and hence a definite prediction requires a precise determination of the GUT scale. 
We leave this issue for future work.

Finally we consider proton decay in our scenario. 
In SUSY GUT models, the proton decay process induced by the dimension-5 operators 
  via color-triplet Higgsino exchanges~\cite{dim5} is severely constrained. 
In evaluating the proton decay amplitude through the dimension-five operators, 
  the information of the fermion Yukawa matrices is necessary.     
Again, we employ the data of the Yukawa matrices obtained in Ref.~\cite{FIM}.  
Recently, in Ref.~\cite{FIM2} the proton decay widths have been calculated in detail with the same Yukawa matrix data 
  and it has been found that the proton lifetime is about 6 times larger than the current 
  experimental limit. 
Here, a special Yukawa structure, namely $f_{11}$ and $f_{12}$ are extremely small, 
  plays a crucial role to suppress the proton decay amplitude. 
Since the same Yukawa matrix data are used, we simply follow Ref.~\cite{FIM2} 
  for our calculation of the proton decay amplitudes. 
The only difference from the calculation in Ref.~\cite{FIM2} is the choice 
  of the particle mass spectrum in our benchmark on Table \ref{table:BM}, 
  while $m_{\tilde q}=\mu=2$ TeV has been taken in Ref.~\cite{FIM2}.  
With our benchmark particle mass spectrum, we have estimated sparticle loop corrections 
  using the formulas given, for example, in Ref.~\cite{Fukuyama:2004xs} 
  and found that the proton decay width is enhanced roughly by a factor of 1.5 
  from the one obtained in Ref.~\cite{FIM2}. 
Therefore, the proton lifetime is about 4 times larger than the current experimental bound. 
This proton lifetime can be tested by the projected Hyper-Kamiokande experiment \cite{Abe:2014oxa}.

\section{Conclusions}
The SUSY minimal SO(10) model is a well motivated GUT scenario for new physics at very high energies, 
  where the SM fermions have Yukawa couplings 
  with only one ${\bf 10}$-plet and one $\overline{\bf 126}$-plet Higgs fields. 
With such a limited number of free parameters, it is highly non-trivial 
  if this model can fit all the data of the SM fermion mass matrices at low energies. 
It has been known that the best fit for all the SM fermion mass matrices 
  is achieved by a $\overline{\bf 126}$-plet Higgs field VEV  
  being at the intermediate scale of around ${\cal O}(10^{13})$ GeV. 
In the minimal SO(10) model with a renormalizable Higgs sector, 
  many exotic states emerge in association with the intermediate-scale SO(10) breaking, 
  so that the success of the SM gauge coupling unification is spoiled.

The fitting of the fermion mass matrices in the minimal SO(10) model has been revisited 
  in Ref.~\cite{FIM} and it has been found that the low-energy fermion mass matrices, 
  except for the down-quark mass predicted to be too low, 
  are very well-fitted without the intermediate scale. 
This result is particularly interesting since the successful gauge coupling unification at the GUT scale is kept intact. 
We have discussed that the too-small down quark mass can be resolved 
  by the SUSY threshold corrections via the gluino loop diagrams 
  with a suitable choice of soft parameters.  
We have found that the flavor-dependent soft parameters are essential 
   in order for the SUSY threshold corrections to work only for resolving the too-small down quark mass.

Motivated by the need of the flavor-dependent soft parameters, we have considered 
   flavor-dependent inputs for the soft parameters at the GUT scale and 
   calculated particle mass spectra at low energies. 
For particle mass spectra, we have taken a variety of phenomenological constraints into account. 
The most important constraints are 
  (i) the SM-like Higgs boson mass of around 125 GeV, 
  (ii) the observed DM relic density for the neutralino LSP and 
  (iii) the discrepancy between the measured value and the SM prediction 
        for the muon anomalous magnetic moment, which is filled by sparticle contributions. 
The resultant sparticle mass spectra are found to be very characteristic such that 
   sleptons in the first and second generations, bino and winos are all light 
   and have masses $< 1$ TeV.  
We present a benchmark particle mass spectrum which satisfies all phenomenological constraints. 
Our scenario with the light sparticles can be tested at the LHC Run-2 in the near future. 
With the benchmark particle mass spectrum, we have estimated the proton lifetime 
  and found it to be about 4 times larger than the current experimental limit, 
  which can be tested in the the projected Hyper-Kamiokande experiment.

\section*{Acknowledgments}
We would like to thank Yukihiro Mimura for useful discussions and comments. 
H.M.T. would like to thank the Department of Physics and Astronomy at the University of Alabama 
   for hospitality during his visit. 
The work of T.F is supported in part by Grant-in-Aid for Science Research 
  from Japan Ministry of Education, Science and Culture (No.~26247036).    
The work of N.O. is supported in part by the United States Department of Energy Grant ($\mbox{DE-SC0013680}$). 
The work of H.M.T. is supported in part by Vietnam Education Foundation, and Vietnam National Foundation for Science and Technology Development (NAFOSTED) 
  under the grant No.~103.01-2014.22.



\begin{thebibliography}{99}

\bibitem{LHC}
https://twiki.cern.ch/twiki/bin/view/AtlasPublic/SupersymmetryPublicResults; 
https://twiki.cern.ch/twiki/bin/view/CMSPublic/PhysicsResultsSUS. 


\bibitem{SMass} 
See, for example, 
  C.~F.~Berger, J.~S.~Gainer, J.~L.~Hewett and T.~G.~Rizzo,
  ``Supersymmetry Without Prejudice,''
  JHEP {\bf 0902}, 023 (2009)
  [arXiv:0812.0980 [hep-ph]]; 
%
J.~A.~Conley, J.~S.~Gainer, J.~L.~Hewett, M.~P.~Le and T.~G.~Rizzo,
  ``Supersymmetry Without Prejudice at the LHC,''
  Eur.\ Phys.\ J.\ C {\bf 71}, 1697 (2011)
  [arXiv:1009.2539 [hep-ph]];  
%
  J.~A.~Conley, J.~S.~Gainer, J.~L.~Hewett, M.~P.~Le and T.~G.~Rizzo,
  ``Supersymmetry Without Prejudice at the 7 TeV LHC,''
  [arXiv:1103.1697 [hep-ph]].



\bibitem{Babu}
  K.~S.~Babu and R.~N.~Mohapatra,
  ``Predictive Neutrino Spectrum In Minimal SO(10) Grand Unification,''
  Phys.\ Rev.\ Lett.\  {\bf 70}, 2845 (1993)
  [hep-ph/9209215].


\bibitem{Matsuda} 
  K.~Matsuda, Y.~Koide and T.~Fukuyama,
  ``Can the SO(10) model with two Higgs doublets reproduce the observed fermion masses?,''
  Phys.\ Rev.\ D {\bf 64}, 053015 (2001)
  [hep-ph/0010026];
%
 K.~Matsuda, Y.~Koide, T.~Fukuyama and H.~Nishiura,
  ``How far can the SO(10) two Higgs model describe the observed neutrino masses and mixings?,''
  Phys.\ Rev.\ D {\bf 65}, 033008 (2002)
  [Erratum-ibid.\ D {\bf 65}, 079904 (2002)]
  [hep-ph/0108202].
  

 \bibitem{FO} 
  T.~Fukuyama and N.~Okada,
  ``Neutrino oscillation data versus minimal supersymmetric SO(10) model,''
  JHEP {\bf 0211}, 011 (2002)
  [hep-ph/0205066].

 
 
\bibitem{Clark}
  T.~E.~Clark, T.~K.~Kuo and N.~Nakagawa,
  ``A SO(10) Supersymmetric Grand Unified Theory,''
  Phys.\ Lett.\  B {\bf 115}, 26 (1982); 
  %
  C.~S.~Aulakh and R.~N.~Mohapatra,
  ``Implications Of Supersymmetric SO(10) Grand Unification,''
  Phys.\ Rev.\  D {\bf 28}, 217 (1983);
  D.~G.~Lee,
  ``Symmetry breaking and mass spectra in the minimal supersymmetric SO(10)
  grand unified theory,''
  Phys.\ Rev. {\bf D49}, 1417 (1994).
  C.~S.~Aulakh, B.~Bajc, A.~Melfo, G.~Senjanovi\'c and F.~Vissani,
  ``The minimal supersymmetric grand unified theory,''
  Phys.\ Lett.\  B {\bf 588}, 196 (2004)
  [arXiv:hep-ph/0306242].
 
 
 
\bibitem{Aulakh} 
T.~Fukuyama, A.~Ilakovac, T.~Kikuchi, S.~Meljanac and N.~Okada,
  ``General formulation for proton decay rate in minimal supersymmetric SO(10) GUT,''
  Eur.\ Phys.\ J.\ C {\bf 42}, 191 (2005)  [hep-ph/0401213];
%
  ``SO(10) group theory for the unified model building,''
  J.\ Math.\ Phys.\  {\bf 46}, 033505 (2005)
  [hep-ph/0405300];  
  ``Higgs masses in the minimal SUSY SO(10) GUT,''
  Phys.\ Rev.\ D {\bf 72}, 051701 (2005)
  [hep-ph/0412348];
%
  B.~Bajc, A.~Melfo, G.~Senjanovic and F.~Vissani,
 ``The Minimal supersymmetric grand unified theory. 1. Symmetry breaking and the particle spectrum,''
 Phys.\ Rev.\ D {\bf 70}, 035007 (2004)
 [hep-ph/0402122]; 
%
  ``Fermion mass relations in a supersymmetric SO(10) theory,''
  Phys.\ Lett.\ B {\bf 634}, 272 (2006)
  [hep-ph/0511352];
 %
  C.~S.~Aulakh and A.~Girdhar,
  ``SO(10) MSGUT: Spectra, couplings and threshold effects,''
  Nucl.\ Phys.\ B {\bf 711}, 275 (2005)
  [hep-ph/0405074].  
      

 \bibitem{Bajc} 
  B.~Bajc, G.~Senjanovic and F.~Vissani,
  ``b - tau unification and large atmospheric mixing: A Case for noncanonical seesaw,''
  Phys.\ Rev.\ Lett.\  {\bf 90}, 051802 (2003)
  [hep-ph/0210207];
  %
  ``Probing the nature of the seesaw in renormalizable SO(10),''
  Phys.\ Rev.\ D {\bf 70}, 093002 (2004)
  [hep-ph/0402140];
%
%
 H.~S.~Goh, R.~N.~Mohapatra and S.~-P.~Ng,
  ``Minimal SUSY SO(10), b tau unification and large neutrino mixings,''
 Phys.\ Lett.\ B {\bf 570}, 215 (2003)
  [hep-ph/0303055];
  %
  ``Minimal SUSY SO(10) model and predictions for neutrino mixings and leptonic CP violation,''
  Phys.\ Rev.\ D {\bf 68}, 115008 (2003)
  [hep-ph/0308197];
%
 B.~Dutta, Y.~Mimura and R.~N.~Mohapatra,
Phys.\ Rev.\ D {\bf 69}, 115014 (2004)
  [hep-ph/0402113];
%
  ``Neutrino masses and mixings in a predictive SO(10) model with CKM CP violation,''
  Phys.\ Lett.\ B {\bf 603}, 35 (2004)
  [hep-ph/0406262].

  

\bibitem{BM}
   K.~S.~Babu and C.~Macesanu,
  ``Neutrino masses and mixings in a minimal SO(10) model,''
  Phys.\ Rev.\ D {\bf 72}, 115003 (2005)
  [hep-ph/0505200].
  
  

\bibitem{Bertolini} 
  S.~Bertolini, T.~Schwetz and M.~Malinsky,
  ``Fermion masses and mixings in SO(10) models and the neutrino challenge to SUSY GUTs,''
  Phys.\ Rev.\ D {\bf 73}, 115012 (2006)
  [hep-ph/0605006].


\bibitem{FIM}
T.~Fukuyama, K.~Ichikawa and Y.~Mimura,
  ``Revisiting fermion mass and mixing fits in the minimal SUSY $SO(10)$ GUT,''
  Phys.\ Rev.\ D {\bf 94}, no. 7, 075018 (2016)
  [arXiv:1508.07078 [hep-ph]].


\bibitem{Camargo-Molina:2013sta}
 J.~F.~Gunion, H.~E.~Haber and M.~Sher,
  ``Charge/Color Breaking Minima and A-Parameter Bounds in Supersymmetric Models,''
  Nucl.\ Phys.\ B {\bf 306}, 1 (1988);
  J.~E.~Camargo-Molina, B.~O'Leary, W.~Porod and F.~Staub,
  ``Stability of the CMSSM against sfermion VEVs,''
  JHEP {\bf 1312}, 103 (2013)
  [arXiv:1309.7212 [hep-ph]].


\bibitem{Kusenko:1996jn} 
  A.~Kusenko, P.~Langacker and G.~Segre,
  ``Phase transitions and vacuum tunneling into charge and color breaking minima in the MSSM,''
  Phys.\ Rev.\ D {\bf 54}, 5824 (1996)
  [hep-ph/9602414].


\bibitem{Hall}
 L.~J.~Hall, R.~Rattazzi and U.~Sarid,
  ``The Top quark mass in supersymmetric SO(10) unification,''
  Phys.\ Rev.\ D {\bf 50}, 7048 (1994)
  [hep-ph/9306309].
  
  
\bibitem{Ibe:2013oha} 
  M.~Ibe, T.~T.~Yanagida and N.~Yokozaki,
  ``Muon g-2 and 125 GeV Higgs in Split-Family Supersymmetry,''
  JHEP {\bf 1308}, 067 (2013)
  [arXiv:1303.6995 [hep-ph]];
  S.~Akula and P.~Nath,
  ``Gluino-driven radiative breaking, Higgs boson mass, muon g-2, and the Higgs diphoton decay in supergravity unification,''
  Phys.\ Rev.\ D {\bf 87}, no. 11, 115022 (2013)
  [arXiv:1304.5526 [hep-ph]];
  M.~A.~Ajaib, I.~Gogoladze, Q.~Shafi and C.~S.~Un,
  ``Split sfermion families, Yukawa unification and muon $g - 2$,''
  JHEP {\bf 1405}, 079 (2014)
  [arXiv:1402.4918 [hep-ph]];
  K.~S.~Babu, I.~Gogoladze, Q.~Shafi and C.~S.~Un,
  ``Muon g-2, 125 GeV Higgs boson, and neutralino dark matter in a flavor symmetry-based MSSM,''
  Phys.\ Rev.\ D {\bf 90}, no. 11, 116002 (2014)
  [arXiv:1406.6965 [hep-ph]];
  M.~Adeel Ajaib, I.~Gogoladze and Q.~Shafi,
  ``GUT-inspired supersymmetric model for h→γγ and the muon g-2,''
  Phys.\ Rev.\ D {\bf 91}, no. 9, 095005 (2015)
  [arXiv:1501.04125 [hep-ph]];
  D.~Chowdhury and N.~Yokozaki,
  ``Muon $g - 2$ in anomaly mediated SUSY breaking,''
  JHEP {\bf 1508}, 111 (2015)
  [arXiv:1505.05153 [hep-ph]];
  I.~Gogoladze, Q.~Shafi and C.~S.~Un,
  ``Reconciling the muon $g-2$, a 125 GeV Higgs boson, and dark matter in gauge mediation models,''
  Phys.\ Rev.\ D {\bf 92}, no. 11, 115014 (2015)
  [arXiv:1509.07906 [hep-ph]];
  N.~Okada and H.~M.~Tran,
  ``125 GeV Higgs boson mass and muon $g-2$ in 5D MSSM,''
  Phys.\ Rev.\ D {\bf 94}, no. 7, 075016 (2016)
  [arXiv:1606.05329 [hep-ph]];
  W.~Yin and N.~Yokozaki,
  ``Splitting mass spectra and muon $g-2$ in Higgs-anomaly mediation,''
  Phys.\ Lett.\ B {\bf 762}, 72 (2016)
  [arXiv:1607.05705 [hep-ph]].
  



\bibitem{typeI} 
  P.~Minkowski,
  ``$\mu \to e\gamma$ at a Rate of One Out of $10^{9}$ Muon Decays?,''
  Phys.\ Lett.\ B {\bf 67}, 421 (1977); 
%
 T.~Yanagida,
  ``Horizontal Symmetry And Masses Of Neutrinos,''
  Conf.\ Proc.\ C {\bf 7902131}, 95 (1979);
%
  M.~Gell-Mann, P.~Ramond and R.~Slansky,
  ``Complex Spinors and Unified Theories,''
  Conf.\ Proc.\ C {\bf 790927}, 315 (1979)
  [arXiv:1306.4669 [hep-th]]; 
%
  S.~L.~Glashow,
  ``The Future of Elementary Particle Physics,''
  NATO Sci.\ Ser.\ B {\bf 61}, 687 (1980); 
%
    R.~N.~Mohapatra and G.~Senjanovic,
    ``Neutrino Masses and Mixings in Gauge Models with Spontaneous Parity Violation,''
    Phys.\ Rev.\ D {\bf 23}, 165 (1981).
    
    
\bibitem{typeII}
 G.~Lazarides, Q.~Shafi and C.~Wetterich,
  ``Proton Lifetime and Fermion Masses in an SO(10) Model,''
  Nucl.\ Phys.\ B {\bf 181}, 287 (1981);
%
  R.~N.~Mohapatra and G.~Senjanovic,
  ``Neutrino Masses and Mixings in Gauge Models with Spontaneous Parity Violation,''
  Phys.\ Rev.\ D {\bf 23}, 165 (1981); 
%
  M.~Magg and C.~Wetterich,
  ``Neutrino Mass Problem and Gauge Hierarchy,''
  Phys.\ Lett.\  {\bf 94B}, 61 (1980); 
%
  J.~Schechter and J.~W.~F.~Valle,
  ``Neutrino Masses in SU(2) x U(1) Theories,''
  Phys.\ Rev.\ D {\bf 22}, 2227 (1980).


\bibitem{Dutta:2005ni} 
 B.~Dutta, Y.~Mimura and R.~N.~Mohapatra,
  ``Neutrino mixing predictions of a minimal SO(10) model with suppressed proton decay,''
  Phys.\ Rev.\ D {\bf 72}, 075009 (2005)
  [hep-ph/0507319].


\bibitem{Bora:2012tx} 
  K.~Bora,
  ``Updated values of running quark and lepton masses at GUT scale in SM, 2HDM and MSSM,''
  Horizon-A Journal of Physics {\bf 2}, 112 (2012)
  [arXiv:1206.5909 [hep-ph]].
   
 
 \bibitem{Agashe:2014kda} 
  K.~A.~Olive {\it et al.}  [Particle Data Group Collaboration],
  ``Review of Particle Physics,''
  Chin.\ Phys.\ C {\bf 38}, 090001 (2014). 
  
  
\bibitem{Capozzi:2013csa} 
  F.~Capozzi, G.~L.~Fogli, E.~Lisi, A.~Marrone, D.~Montanino and A.~Palazzo,
  Phys.\ Rev.\ D {\bf 89}, 093018 (2014)
  [arXiv:1312.2878 [hep-ph]].


\bibitem{FIM2}
  T.~Fukuyama, K.~Ichikawa and Y.~Mimura,
  ``Relation between proton decay and PMNS phase in the minimal SUSY $SO(10)$ GUT,''
  Phys.\ Lett.\ B {\bf 764}, 114 (2017)
  [arXiv:1609.08640 [hep-ph]].
  
    
\bibitem{Allanach:2001kg} 
  B.~C.~Allanach,
  ``SOFTSUSY: a program for calculating supersymmetric spectra,''
  Comput.\ Phys.\ Commun.\  {\bf 143}, 305 (2002)
  [hep-ph/0104145].
  

\bibitem{Belanger:2014vza} 
  G.~Belanger, F.~Boudjema, A.~Pukhov and A.~Semenov,
  ``micrOMEGAs4.1: two dark matter candidates,''
  Comput.\ Phys.\ Commun.\  {\bf 192}, 322 (2015)
  [arXiv:1407.6129 [hep-ph]].


\bibitem{Belanger:2004yn} 
  G.~Belanger, F.~Boudjema, A.~Pukhov and A.~Semenov,
  ``micrOMEGAs: Version 1.3,''
  Comput.\ Phys.\ Commun.\  {\bf 174}, 577 (2006)
  [hep-ph/0405253].

\bibitem{Belanger:2001fz} 
  G.~Belanger, F.~Boudjema, A.~Pukhov and A.~Semenov,
  ``MicrOMEGAs: A Program for calculating the relic density in the MSSM,''
  Comput.\ Phys.\ Commun.\  {\bf 149}, 103 (2002)
  [hep-ph/0112278].


\bibitem{Bechtle:2008jh} 
  P.~Bechtle, O.~Brein, S.~Heinemeyer, G.~Weiglein and K.~E.~Williams,
  ``HiggsBounds: Confronting Arbitrary Higgs Sectors with Exclusion Bounds from LEP and the Tevatron,''
  Comput.\ Phys.\ Commun.\  {\bf 181}, 138 (2010)
  [arXiv:0811.4169 [hep-ph]].

\bibitem{Bechtle:2011sb} 
  P.~Bechtle, O.~Brein, S.~Heinemeyer, G.~Weiglein and K.~E.~Williams,
  ``HiggsBounds 2.0.0: Confronting Neutral and Charged Higgs Sector Predictions with Exclusion Bounds from LEP and the Tevatron,''
  Comput.\ Phys.\ Commun.\  {\bf 182}, 2605 (2011)
  [arXiv:1102.1898 [hep-ph]].

\bibitem{Bechtle:2013gu} 
  P.~Bechtle, O.~Brein, S.~Heinemeyer, O.~Stal, T.~Stefaniak, G.~Weiglein and K.~Williams,
  ``Recent Developments in HiggsBounds and a Preview of HiggsSignals,''
  PoS CHARGED {\bf 2012}, 024 (2012)
  [arXiv:1301.2345 [hep-ph]].

\bibitem{Bechtle:2013wla} 
  P.~Bechtle, O.~Brein, S.~Heinemeyer, O.~Stal, T.~Stefaniak, G.~Weiglein and K.~E.~Williams,
  ``$\mathsf{HiggsBounds}-4$: Improved Tests of Extended Higgs Sectors against Exclusion Bounds from LEP, the Tevatron and the LHC,''
  Eur.\ Phys.\ J.\ C {\bf 74}, no. 3, 2693 (2014)
  [arXiv:1311.0055 [hep-ph]].

\bibitem{Kraml:2013mwa} 
  S.~Kraml, S.~Kulkarni, U.~Laa, A.~Lessa, W.~Magerl, D.~Proschofsky-Spindler and W.~Waltenberger,
  ``SModelS: a tool for interpreting simplified-model results from the LHC and its application to supersymmetry,''
  Eur.\ Phys.\ J.\ C {\bf 74}, 2868 (2014)
  [arXiv:1312.4175 [hep-ph]].



\bibitem{Aad:2015zhl} 
  G.~Aad {\it et al.} [ATLAS and CMS Collaborations],
  ``Combined Measurement of the Higgs Boson Mass in $pp$ Collisions at $\sqrt{s}=7$ and 8 TeV with the ATLAS and CMS Experiments,''
  Phys.\ Rev.\ Lett.\  {\bf 114}, 191803 (2015)
  [arXiv:1503.07589 [hep-ex]].


\bibitem{Bennett:2006fi} 
  G.~W.~Bennett {\it et al.} [Muon g-2 Collaboration],
  ``Final Report of the Muon E821 Anomalous Magnetic Moment Measurement at BNL,''
  Phys.\ Rev.\ D {\bf 73}, 072003 (2006)
  [hep-ex/0602035].



\bibitem{Roberts:2010cj} 
  B.~L.~Roberts,
  ``Status of the Fermilab Muon $(g-2)$ Experiment,''
  Chin.\ Phys.\ C {\bf 34}, 741 (2010)
  [arXiv:1001.2898 [hep-ex]].


\bibitem{Hagiwara:2011af} 
  K.~Hagiwara, R.~Liao, A.~D.~Martin, D.~Nomura and T.~Teubner,
  ``$(g-2)_\mu$ and $\alpha(M_Z^2)$ re-evaluated using new precise data,''
  J.\ Phys.\ G {\bf 38}, 085003 (2011)
  [arXiv:1105.3149 [hep-ph]].


\bibitem{Aoyama:2012wk} 
  T.~Aoyama, M.~Hayakawa, T.~Kinoshita and M.~Nio,
  ``Complete Tenth-Order QED Contribution to the Muon g-2,''
  Phys.\ Rev.\ Lett.\  {\bf 109}, 111808 (2012)
  [arXiv:1205.5370 [hep-ph]].


\bibitem{hfag}
  Y.~Amhis {\it et al.} [Heavy Flavor Averaging Group Collaboration],
  ``Averages of B-Hadron, C-Hadron, and tau-lepton properties as of early 2012,''
  arXiv:1207.1158 [hep-ex].
  


\bibitem{Olive:2016xmw} 
  C.~Patrignani {\it et al.} [Particle Data Group Collaboration],
  Chin.\ Phys.\ C {\bf 40}, no. 10, 100001 (2016).



\bibitem{Antonelli:2008jg} 
  M.~Antonelli {\it et al.} [FlaviaNet Working Group on Kaon Decays Collaboration],
  ``Precision tests of the Standard Model with leptonic and semileptonic kaon decays,''
  arXiv:0801.1817 [hep-ph].




\bibitem{Ade:2015xua} 
  P.~A.~R.~Ade {\it et al.} [Planck Collaboration],
  ``Planck 2015 results. XIII. Cosmological parameters,''
  arXiv:1502.01589 [astro-ph.CO].


\bibitem{Akerib:2015rjg} 
  D.~S.~Akerib {\it et al.} [LUX Collaboration],
  ``Improved Limits on Scattering of Weakly Interacting Massive Particles from Reanalysis of 2013 LUX Data,''
  Phys.\ Rev.\ Lett.\  {\bf 116}, no. 16, 161301 (2016)
  [arXiv:1512.03506 [astro-ph.CO]]; 
  D.~S.~Akerib {\it et al.},
  ``Results from a search for dark matter in the complete LUX exposure,''
  arXiv:1608.07648 [astro-ph.CO].
  
  
\bibitem{Aartsen:2016exj} 
  M.~G.~Aartsen {\it et al.} [IceCube Collaboration],
  ``Improved limits on dark matter annihilation in the Sun with the 79-string IceCube detector and implications for supersymmetry,''
  JCAP {\bf 1604}, no. 04, 022 (2016)
  [arXiv:1601.00653 [hep-ph]].


\bibitem{Moroi} 
  T.~Moroi,
  ``The Muon anomalous magnetic dipole moment in the minimal supersymmetric standard model,''
  Phys.\ Rev.\ D {\bf 53}, 6565 (1996)
  Erratum: [Phys.\ Rev.\ D {\bf 56}, 4424 (1997)]
  [hep-ph/9512396].


\bibitem{Hisano-etal}
  J.~Hisano, T.~Moroi, K.~Tobe, M.~Yamaguchi and T.~Yanagida,
  ``Lepton flavor violation in the supersymmetric standard model with seesaw induced neutrino masses,''
  Phys.\ Lett.\ B {\bf 357}, 579 (1995)
  [hep-ph/9501407]; 
%
 J.~Hisano, T.~Moroi, K.~Tobe and M.~Yamaguchi,
  ``Lepton flavor violation via right-handed neutrino Yukawa couplings in supersymmetric standard model,''
  Phys.\ Rev.\ D {\bf 53}, 2442 (1996)
  [hep-ph/9510309].


\bibitem{MEG}
  T.~Mori [MEG Collaboration],
  ``Final Results of the MEG Experiment,''
  arXiv:1606.08168 [hep-ex].
  
  
  
\bibitem{dim5}
S.~Weinberg,
  ``Supersymmetry at Ordinary Energies. 1. Masses and Conservation Laws,''
  Phys.\ Rev.\ D {\bf 26}, 287 (1982);  
%
N.~Sakai and T.~Yanagida,
  ``Proton Decay in a Class of Supersymmetric Grand Unified Models,''
  Nucl.\ Phys.\ B {\bf 197}, 533 (1982).


  
\bibitem{Fukuyama:2004xs} 
  T.~Fukuyama, A.~Ilakovac, T.~Kikuchi, S.~Meljanac and N.~Okada,
  ``General formulation for proton decay rate in minimal supersymmetric SO(10) GUT,''
  Eur.\ Phys.\ J.\ C {\bf 42}, 191 (2005)
  [hep-ph/0401213].



\bibitem{Abe:2014oxa} 
  K.~Abe {\it et al.} [Hyper-Kamiokande Working Group],
  ``A Long Baseline Neutrino Oscillation Experiment Using J-PARC Neutrino Beam and Hyper-Kamiokande,''
  arXiv:1412.4673 [physics.ins-det];
  K.~Abe {\it et al.} [Hyper-Kamiokande Proto- Collaboration],
  ``Physics potential of a long-baseline neutrino oscillation experiment using a J-PARC neutrino beam and Hyper-Kamiokande,''
  PTEP {\bf 2015}, 053C02 (2015)
  doi:10.1093/ptep/ptv061
  [arXiv:1502.05199 [hep-ex]]. 
  K.~Abe {\it et al.} [Hyper-Kamiokande proto- Collaboration],
  ``Physics Potentials with the Second Hyper-Kamiokande Detector in Korea,''
  arXiv:1611.06118 [hep-ex].
  
  
  

\end{thebibliography}
\end{document}